\documentclass[
 reprint,
 amsmath,amssymb,
 aps,prb,eps,superscriptaddress,10pt, twocolumn
]{revtex4}
\usepackage{gensymb}
\usepackage{graphicx}
\usepackage{dcolumn}
\usepackage{bm}
\usepackage{xcolor}
\usepackage{appendix}
\usepackage{inputenc}
\usepackage{changes}
\definechangesauthor[name=Markus Aichhorn, color=red]{MA}

\begin{document}

\preprint{APS/123-QED}


\title{Effect of geometry on magnetism of Hund's metals: A case study with BaRuO$_3$}

\author{Hrishit Banerjee}
\affiliation{Yusuf Hamied Department of Chemistry, University of Cambridge, Lensfield Road, CB2 1EW, Cambridge, UK}
\affiliation{Institute of Theoretical and Computational Physics, Graz University of Technology, NAWI Graz, Petersga{\ss}e 16, Graz, 8010, Austria.
}
\author{Hermann Schnait}
\author{Markus Aichhorn}
\affiliation{Institute of Theoretical and Computational Physics, Graz University of Technology, NAWI Graz, Petersga{\ss}e 16, Graz, 8010, Austria.
}
\author{Tanusri Saha-Dasgupta}
\email{t.sahadasgupta@gmail.com}
\affiliation{S. N. Bose National Centre for Basic Sciences, Block JD, Sector - III, Salt lake, Kolkata-700106, India.}

\begin{abstract}
In order to explore the effects of structural geometry on properties of correlated metals we investigate the magnetic properties of cubic (3C) and hexagonal (4H) BaRuO$_3$. While the 3C variant of BaRuO$_3$ is ferromagnetic below 60\,K, the 4H phase does not show any long-range magnetic order, however, there is experimental evidence of short-range antiferromagnetic correlations.
  Employing a combination of computational tools, namely density-functional theory and dynamical mean-field theory calculations, 
  we probe the origin of contrasting magnetic
  properties of BaRuO$_3$ in the 3C and 4H structures. Our study reveals that the difference in connectivity of RuO$_6$ octahedra in the two phases results in different Ru-O covalency, which in turn influences substantially the strengths of screened interaction values for Hubbard $U$ and Hund's rule $J$. With estimated $U$ and $J$ values, the 3C phase turns out to be a ferromagnetic metal,
  while the 4H phase shows paramagnetic behavior with vanishing ordered moments. However, this paramagnetic phase bears signatures of antiferromagnetic correlations, as confirmed
  by a calculation of the magnetic susceptibility.
  We find that the 4H phase is found to be at the verge of antiferromagnetic long-range order, which can be
  stabilized upon slight changes of screened Coulomb parameters $U$ and $J$, opening up the possibility of achieving a rare example of an antiferromagnetic metal.
\end{abstract}
\maketitle

\section{Introduction}

Transition metal oxides (TMO) represent a class of compounds exhibiting a plethora of fascinating physical properties.\cite{khomskii_2014} for instance, the intriguing interplay of charge, spin, and orbital degrees of freedom in the context of srong correlations 
opens up a scientifically rewarding playground. While most studies focus on this interplay of charge, spin and orbitals, the effect of geometry of the underlying structure is comparably less explored. This is, however, an equally important issue, given the fact that keeping the basic motif of transition metal-oxygen octahedra, the connectivity of these octahedra in transition metal oxides can greatly vary from compound to compound.  

To explore the influence of variation of connectivity in a systematic manner, it is desirable to find a structural variation within the same chemical composition in systems where the interplay of geometry and correlation effect is expected to be strong. Moving down the periodic table from the 3$d$ to the 4$d$ transition-metal series, the covalency between transition metal and oxygen increases. This leads to a large crystal field splitting between e$_g$ and t$_{2g}$ states that is usually of the order of the local screened Coulomb interaction $U$. Furthermore, the larger spatial extent of the 4$d$ orbitals produce a larger band width as compared to 3$d$ materials. 
As a result, the 4$d$ compounds prefer a low-spin state rather than the high-spin state, with occupied t$_{2g}$ and empty e$_g$ configurations. While 4$d$ compounds generally show a smaller screened Coulomb interaction $U$ as compared to 3$d$ compounds, the multi-orbital nature of the problem in the t$_{2g}$ manifold makes the Hund's coupling $J$ an important parameter.\cite{Georges2013} From that perspective, the interplay between the structural aspects and the electronic correlations driven by $U$ and $J$ in 4$d$ TMOs may turn out to  be more interesting than that for their 3$d$ counterparts. 

Ruthenates are an ideal playground for studying the 4$d$ physics described above. A large number of ruthenate compounds have been experimentally synthesized, and a number of interesting physical phenomena
has been reported, most of them being related to strong correlation effects. Sr$_2$RuO$_4$ showing unconventional $p$-wave superconductivity,\cite{Maeno1994} SrRuO$_3$ showing high Curie temperatures under compression as well as sensitivity of longitudinal resistivity and magnetic anisotropy to differential methods of strain application,\cite{Daisuke2013} and BaRu$_6$O$_{12}$ showing a quantum phase transition in transport and magnetic properties are just a few examples to mention. 

Among the ruthenate TMOs, BaRuO$_3$ (BRO), which is the sister compound of the well studied compounds SrRuO$_3$ and CaRuO$_3$, offers a perfect platform for the exploration of the above mentioned geometry effects in a systematic manner. Both SrRuO$_3$ and BRO, which are isoelectronic ruthenates
(Ru$^{4+}$ with 4d$^4$ electronic configuration), have been considered as prototypical examples of Hund’s metals with nearly spin-frozen states.\cite{Millis2016, Dasari2016} However, BRO is reported to exhibit structural and physical properties different from SrRuO$_3$ or CaRuO$_3$. The
presence of Ba$^{2+}$ at A site, which has a larger ionic size compared to Sr$^{2+}$ or Ca$^{2+}$, leads to a tolerance factor of BRO of
$t = \frac{(r_{Ba} +r_{O})}{\sqrt{2} (r_{Ru} + r_O )} > 1$ (where $r_{Ba}$, $r_{Ru}$ and $r_O$ correspond to the radius of
Ba, Ru and O ions, respectively), favoring the hexagonal polytype as opposed to SrRuO$_3$ or CaRuO$_3$, for which $t< 1$
favors an orthorhombic structure with cubic stacking. Depending on the synthesis pressure, a sequence of structural types in BRO
is reported from 9R (ambient pressure) to 4H (3\,GPa) to 6H (5\,GPa), all based on hexagonal symmetry.\cite{ZHAO2007} Interesting enough,
the cubic 3C phase of BRO could also be stabilized under very high pressure conditions,\cite{Jin2008} giving rise to unique opportunity
of studying the influence of hexagonal versus cubic connectivity of RuO$_6$ octahedra within the same chemical formula of BRO.

In this study, we take up the 3C and 4H phase of BRO as case study to investigate the effect of geometry on correlation-driven magnetism and electronic structure of 4$d$ TMOs. Cubic BRO with perovskite structure and corner-shared RuO$_6$ octahedra is a ferromagnetic metal, with a $T_c\sim 60$\,K.\cite{Jin2008} On the other hand, the moderate-pressure 4H phase, which exhibits non-perovskite hexagonal geometry with face-shared dimers of RuO$_6$ octahedra does not seem to order magnetically in experimental studies, although some signatures of anti-ferromagnetic correlations have been reported,\cite{Gulino1995} suggesting a paramagnetic metal as a ground state.\cite{ZHAO2007} The change in geometry from cubic to hexagonal, thus, appears to have a profound effect on the material's properties. We investigate this issue by a combination of a variety of tools, ranging from first-principles density-functional theory calculations to constrained random-phase approximation and dynamical mean-field theory. For the latter, we employ both a continuous-time quantum Monte Carlo technique using the hybridisation expansion, as well as the fork tensor product states method. Our study reveals a Hund’s metallic state for both 3C and 4H phases, reflected in a transition from a generalized Fermi liquid to non-Fermi liquid behaviour upon variation of Hund’s rule coupling. This emphasizes the importance of correlation effects for the description of the properties of BRO in both 3C and 4H phases. Ferromagnetically ordered phases, short-range antiferromagnetic fluctuations, and long-range ordered antiferromagnetic phases are found for both 3C and 4H in paramater space of Coulomb correlation $U$ and Hund’s coupling $J$. The central finding of our study is that the change in connectivity between 3C and 4H results in a change in metal-oxygen hybridization which influences the electronic screening, amounting distinctly different estimated values of $U$ and $J$ for 3C and 4H, resp. This difference in the parameters places the 3C phase in the ferromagnetically ordered regime, and 4H in in the paramagnetic regime with short-range antiferromagnetic correlations. It is intriguing to note that a slight variation of the $U$ parameter would stablize a rare and exotic example of an antiferromagnetically ordered metallic phase in the 4H variant, which could be used in spintronics applications. 

\section{Computational details}
Our density-functional theory (DFT) calculations for structural relaxation were carried out in a plane-wave basis with projector-augmented wave (PAW) potentials~\cite{blochl} as implemented in the Vienna Ab-initio Simulation Package
(VASP).\cite{kresse, kresse01}
For ionic relaxations using the VASP package, internal positions of the atoms were allowed to relax until the forces became less than 0.005\,eV/\AA. An energy cutoff of 550\,eV, and a 6$\times$6$\times$4 Monkhorst–Pack $k$-points mesh provided good convergence of the total energy.
Our ab-initio dynamical mean-field theory (DMFT) calculations are based on the full-potential augmented plane-wave basis as implemented in
\textsc{wien2k}~\cite{wien2k}. 
For these calculations, we used the largest possible muffin-tin radii, and the
basis set plane-wave cutoff as defined by
${R_{\text{min}}\!\cdot\!K_{\text{max}}}=7.5$, where $R_{\text{min}}$ is the
muffin-tin radius of the oxygen atoms.
In all our DFT calculations, we chose as
exchange-correlation functional the generalized gradient approximation (GGA), implemented following the Perdew Burke Ernzerhof (PBE) prescription~\cite{pbe}.
The consistency between VASP and \textsc{wien2k} results have been cross-checked. 

We also perform constrained Random Phase Approximation (cRPA) calculations within VASP, with the states of interest dervived from a Wannier90 projection method,\cite{Pizzi2020} to have an estimate of the Hubbard $U$ and Hund's $J$ values for both geometries. This involves a three-step procedure: a DFT groundstate calculation, a calculation to obtain a number of virtual orbitals, and the actual cRPA calculation. 
For the Wannier projections an energy window from $-3$\,eV to $+2$\,eV around the Fermi energy was chosen, and  projections were done to the t$_{2g}$ states. 
A large number of bands (96 bands) were taken in account for the G0W0 calculation. 
The screened Hubbard $U$ and Hund's $J$ are
obtained from the calculation as the static $\omega=0$ limit of the frequency-dependent cRPA interactions.

We perform the DMFT calculations in a basis set of projective Wannier functions, which were calculated using the TRIQS/DFTTools package~\cite{aichhorn1, aichhorn2, aichhorn3}  based on the TRIQS libraries\cite{triqs}. For both paramagnetic and magnetic calculations, only Ru t$_{2g}$ orbitals have been considered for the DMFT calculation, since the e$_g$ orbitals are higher in energy and are, thus, empty.
The Anderson impurity problems were solved using the
continuous-time quantum Monte Carlo algorithm in the hybridization expansion
(CT-HYB)\cite{werner06} as implemented in the
TRIQS/CTHYB package.\cite{pseth-cpc} We
performed one-shot calculations, with the double-counting correction treated in the fully-localized
limit.\cite{anisimov93} We used the density-density variant of the Kanamori interaction.\cite{kanamori63}
For our calculations we did not only use the cRPA values but also $U$ values ranging from 1.3-3\,eV and $J$ values ranging from 0.3-0.5\,eV to explore the phase diagram. 
We set the intra-orbital interaction to be 
$U^{'} =U- 2J$. 
Real-frequency results have been obtained using the maximum-entropy method of analytic continuation as implemented in the TRIQS/MAXENT application.\cite{maxent}

As quantum Monte Carlo solvers are limited to higher temperatures, we also employed an impurity solver based on matrix product states in a special geometry, the fork-tensor-product-states (FTPS) solver.\cite{ftps}
This allows efficient $T=0$ calculations for multi-orbital systems directly on the real frequency axis.
To do so, we discretized the hybridization function using 50 bath-sites per spin, calculated the ground-state using a density-matrix-renormalization group algorithm. The time-evolution for calculating the interacting Greens functions is done using the time-dependent variational principle using 100 time-steps of length $dt=0.1$.\cite{daniel2020}
The calculations were checked for consistency with the full Kanamori Hamiltonian for the 3C structure.

\begin{figure}
    \centering
    \includegraphics[width=\columnwidth]{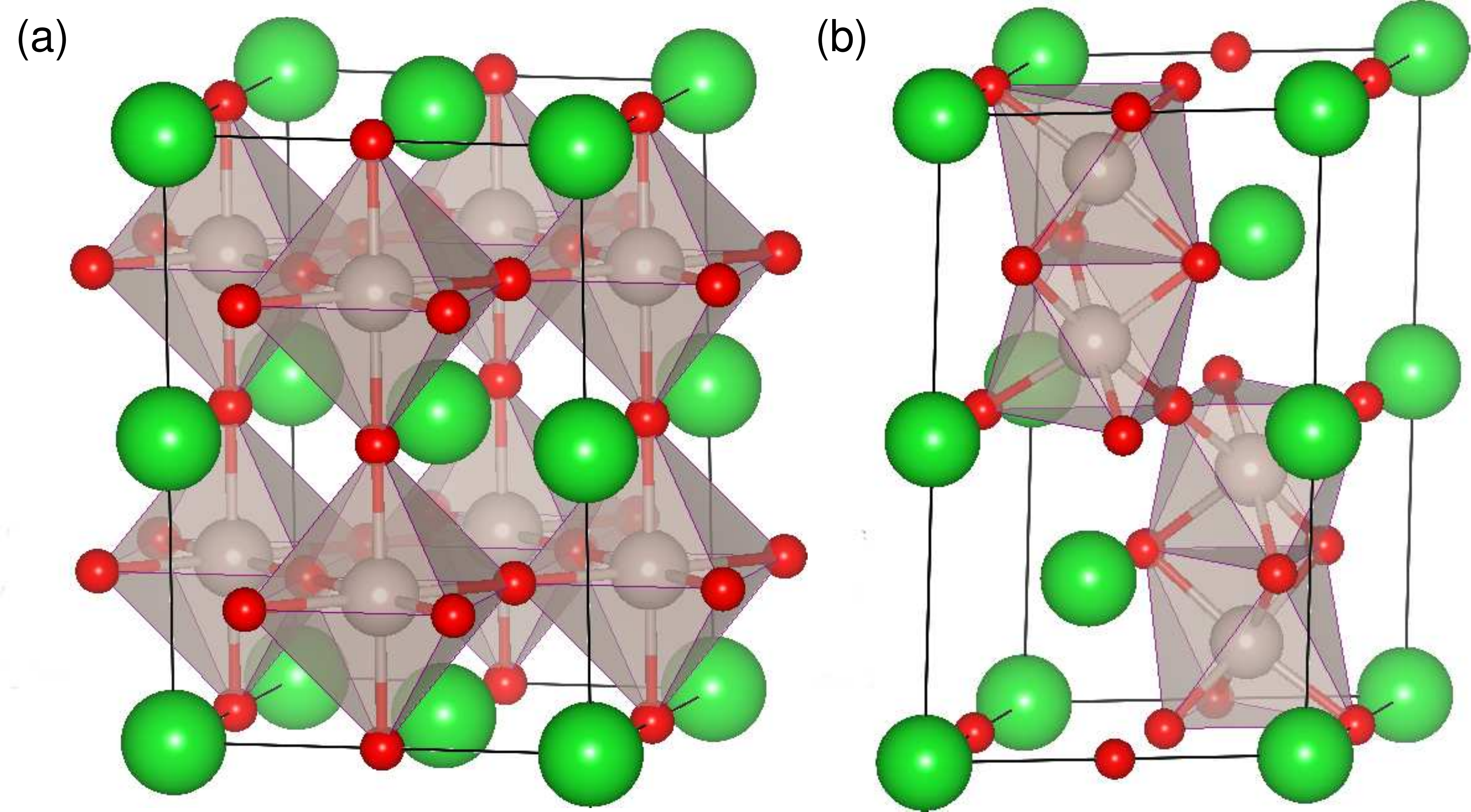}
    \caption{Structures of the cubic 3C and hexagonal 4H phase of BRO. The cubic phase is a perovskite structure with all corner-sharing RuO$_6$ octahedra, while the 4H hexagonal phase consists of face-shared dimers of RuO$_6$ octahedra, corner-sharing with each other.}
    \label{struct}
\end{figure}

\section{Crystal Structure and DFT band structure}

In Fig.~\ref{struct} we show the crystal structure of both the 3C and 4H phase. As already mentioned, depending on the synthesis condition,
the stacking of BaO$_3$ layers can be of the hexagonal close-packed stacking ($h$) or the cubic close-packed stacking ($c$). These two
different stacking patterns give rise to face-shared and corner-shared connectivity of neighboring RuO$_6$ octahedra in 4H and 3C phases,
respectively.

The 3C structure has a perfect cubic symmetry Pm$\bar{3}$m, and has lattice constants of $a=b=c=4.075$\,\AA, and unit cell $\alpha=\beta=\gamma=90\degree$. \cite{Jin2008}
 The RuO$_6$ octahedra are all corner sharing, with the corner-shared Ru-O-Ru angle of 180 $\degree$, as in perfect cubic symmetry.

The 4H structure has a hexagonal symmetry P6$_3$/mmc, and has lattice constants of $a=b=5.729$\,\AA{} and $c=9.5$\,\AA, and unit cell angles $\alpha=\beta=90\degree$ and $\gamma=120\degree$, as given by hexagonal symmetry. The RuO$_6$ octahedra form face-shared dimers, and the dimers are corner-shared between themselves, giving
rise to a $hchc$ stacking sequence, which indicates a stacking of alternate hexagonal and cubic units. The face-shared dimers have an internal Ru-O-Ru angle of $78.5\degree$, while the corner shared dimers have a Ru-O-Ru angle of $180\degree$.\cite{HONG1997}
The percentage of corner-shared connectivity is $100\%$ and $50\%$ for
the 3C and 4H phases, respectively. This difference in connectivity plays a big role in the properties of the two phases
as we shall see in a later section.

\begin{figure}
    \centering
    \includegraphics[width=\columnwidth]{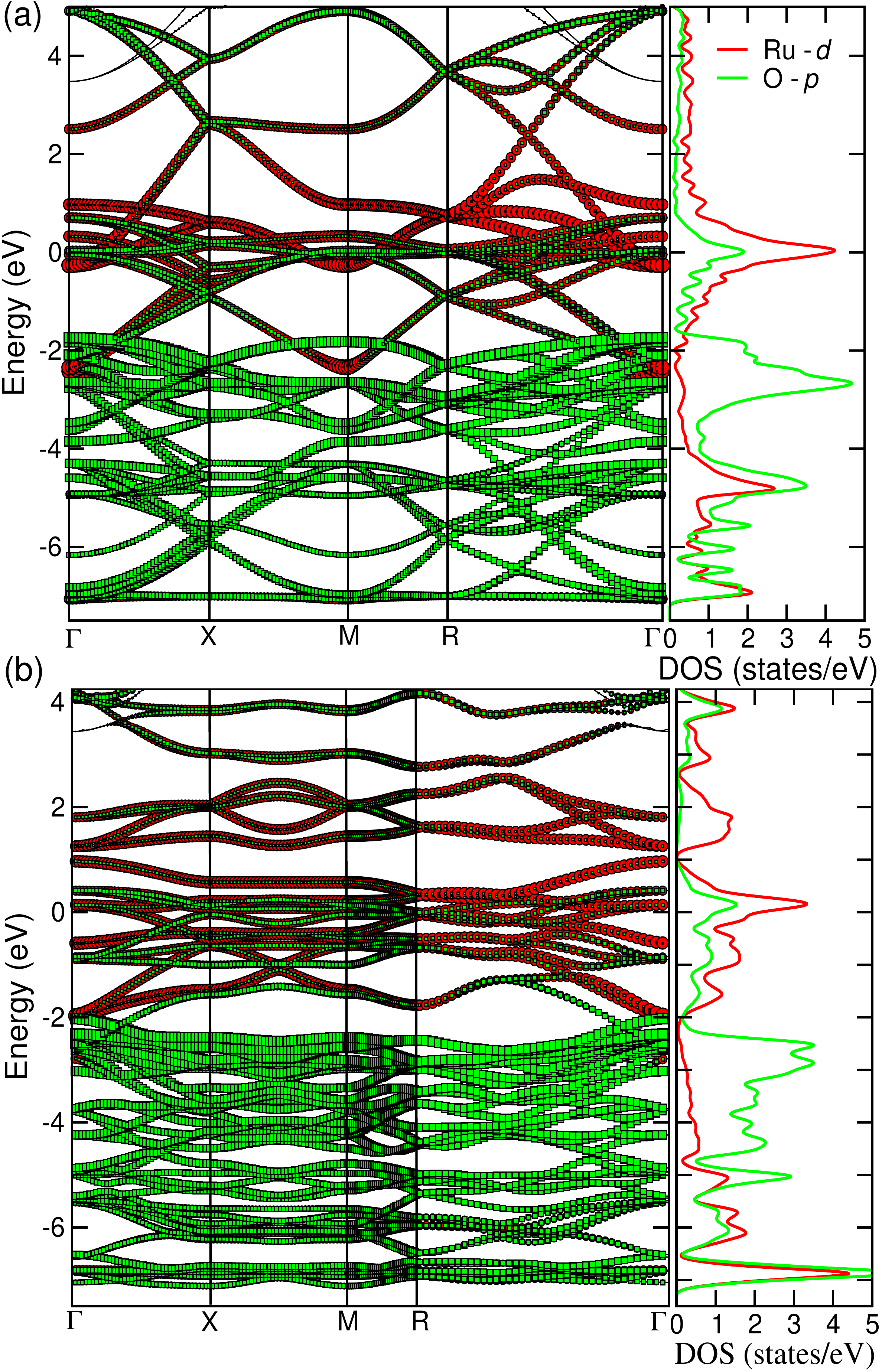}
    \caption{Non-magnetic DFT band structure of the (a) cubic 3C and (b) hexagonal 4H phase of BRO, projected to Ru $d$ and O $p$ states. }
    \label{dftbands}
\end{figure}

In Fig.~\ref{dftbands}, we present the DFT band structure and density of states (DOS) obtained from non-magnetic plane-wave calculations using VASP of 3C and 4H phases.
The states are projected to Ru $d$ (red symbols and lines in Fig.~\ref{dftbands}) and O $p$ orbital degrees of freedom (green symbols and lines in Fig.~\ref{dftbands}). Within the non-spin-polarized scheme of calculations, both systems are metallic with
Ru t$_{2g}$ states crossing the Fermi level, with a strong admixture of O $p$ states. The computed electronic structure is in good agreement with what has been reported
earlier.\cite{Kanungo_2013} Comparing the DFT electronic structure of 3C and 4H we observe a marked difference, in terms of Ru $d$ band width as well
as the charge-transfer energy between Ru $d$ and O $p$ states, which we infer from the
band centers of Ru $d$ and O $p$ bands. The latter is expected to be important in the calculation of the screening of Coulomb parameters $U$ and $J$. We further notice multiple peaks in the DOS of 4H compared to that of 3C, which arises due to bonding-antibonding splitting of trigonally split Ru t$_{2g}$
states as discussed in Ref.~\onlinecite{Kanungo_2013}. It has also been seen in previous calculations that there is an energy gap of approximately 3\,eV between the t$_{2g}$ levels around Fermi energy and the e$_g$ levels at higher energies.\cite{Kanungo_2013} Moreover, Wannier projections to both the t$_{2g}$ subspace as well as to the full $d$ manifold 
yields an occupation of four electrons. Hence, only the t$_{2g}$ levels 
are relevant in our case and are considered in the subsequent DMFT study of the low-spin state of Ru.

\begin{figure}
    \centering
    \includegraphics[width=\columnwidth]{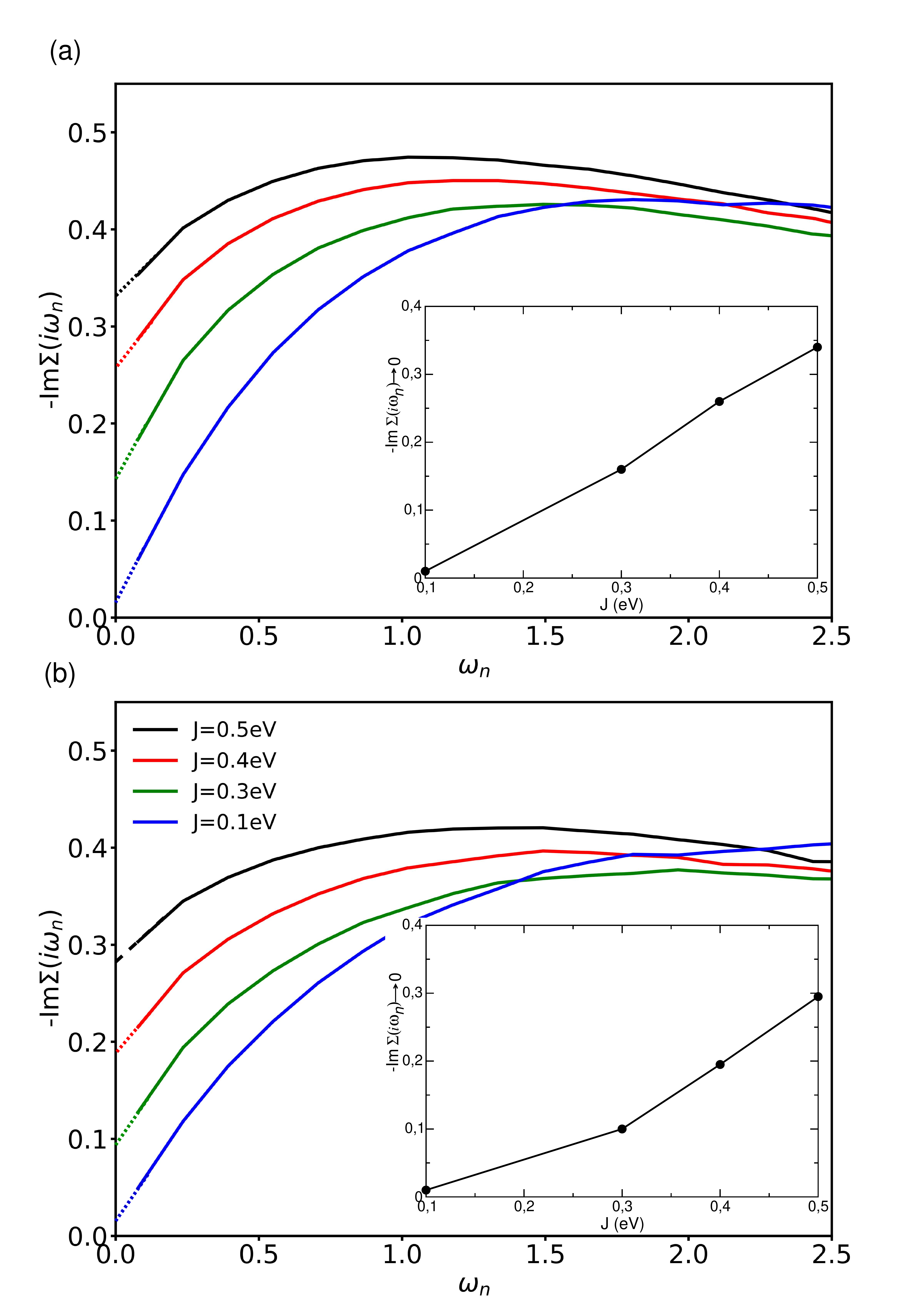}
    \caption{The Imaginary part of self energies of the (a) cubic 3C and (b) hexagonal 4H phase of BRO. Both the phases show a transition from a generalised Fermi liquid to a non Fermi liquid behaviour, depending on the choice of Hund $J$ parameter for a fixed $U=2.3$\,eV. The Inset shows the $y$-axis intercepts obtained by extrapolation as a function of the value of $J$. The extrapolation of the imaginary part of the self energies to $\omega_n=0$ is shown by dotted lines.}
    \label{selfen}
\end{figure}

\section{Single-impurity Dynamics at Finite Temperature}

To take into account the correlation effect of Ru 4$d$ states, we first carry out single-impurity DMFT calculations for Ru t$_{2g}$ based low-energy Hamiltonians defined in the basis of DFT-derived Wannier functions. A local Coulomb
interaction of density-density type between the orbitals is introduced. The interaction part of the Hamiltonian 
is given by,
\begin{align*}
  H_{ii}^{int} &= \sum_{i,m=1}^{3} U n_{i,m\uparrow} n_{i,m\downarrow} \\
  &+ \sum_{i,m \ne m^{'}} \sum_{\sigma,\sigma^{'}} (V - J\delta_{\sigma,\sigma^{'}}) n_{i,m\sigma} n_{i,m^{'}\sigma^{'}},
\end{align*}
where $i$ is the lattice site, and $m$ and $m^{'}$ represent orbital indices. $U$ is the Coulomb repulsion between two electrons
with opposite spin in the same orbital. Orbital rotational symmetry is imposed by setting $V = U - 2J$, where $J$ is the
Hund’s coupling, which lowers the energy of a configuration with different orbitals ($m \ne m^{'}$), and parallel
spins $\sigma = \sigma^{'}$. In this section, the effective impurity problem is solved within DMFT by using the hybridization expansion continuous-time quantum
Monte Carlo which works at finite temperatures. 
In the following, we vary the value of $U$  
within a range of 1-4\,eV and $J$ in the range of 0.1-0.5\,eV, which are sensible parameter ranges for 4$d$ transition metal oxides.\cite{Dasari2016}

\subsection{Generalized Fermi Liquid to Non Fermi Liquid Crossover}

A crossover from a generalised Fermi liquid to a non Fermi liquid behaviour with changes of Hund coupling $J$ has been demonstrated for the 3C phase in a previous study.\cite{Dasari2016} 
We find the same to be true for the 4H phase, as determined by the single-impurity DMFT self-energy $\Sigma (i\omega_n)$. Fig.~\ref{selfen} summarizes the
results which shows the imaginary part of Matsubara self-energies of 3C and 4H for fixed $U = 2.3$\,eV and inverse temperature $\beta = 40 eV^{-1}$, for a range of $J$ values. For small values of $J$, the
low-frequency behavior of the self-energy has a generalized Fermi liquid (GFL) behavior, given by $-\textrm{Im} \Sigma (i \omega_n)\sim a \omega_n^{\alpha}$, where
0 $\le \alpha \le$ 1. Upon increasing $J$, a deviation from such a power-law behavior is found at low $\omega_n$, manifested as non-zero intercept of $-\textrm{Im}\Sigma (i\omega_n)$,
which is characteristic of non-Fermi liquid (NFL) behavior. For 3C (Fig.~\ref{selfen} (a)), we observe a GFL behaviour for $J=0.1$\,eV, with onset of NFL behaviour at $J=0.3$\,eV
which increases progressively to $J=0.5$\,eV. A very similar behaviour is observed for 4H (Fig.~\ref{selfen} (b)), which confirms to importance of Hund's coupling for strong-correlation effects
in both 3C and 4H.

It is to be noted here that at finite $T$ there will always be some small finite intercept of the imaginary part of self energy, since $-\textrm{Im}\Sigma(\omega=0)$ varies as $T^2$, as we also see in case of $J=0.1$\,eV. However, by extrapolation one can see that this value is negligibly small. 
As function of temperature, this behavior has also been coined coherence-incoherence crossover, when the scattering grows much faster as $T^2$ would suggest.\cite{mravlje2011}


\subsection{Correlation driven Magnetism and Electronic structure}

\begin{figure}
    \centering
    \includegraphics[width=\columnwidth]{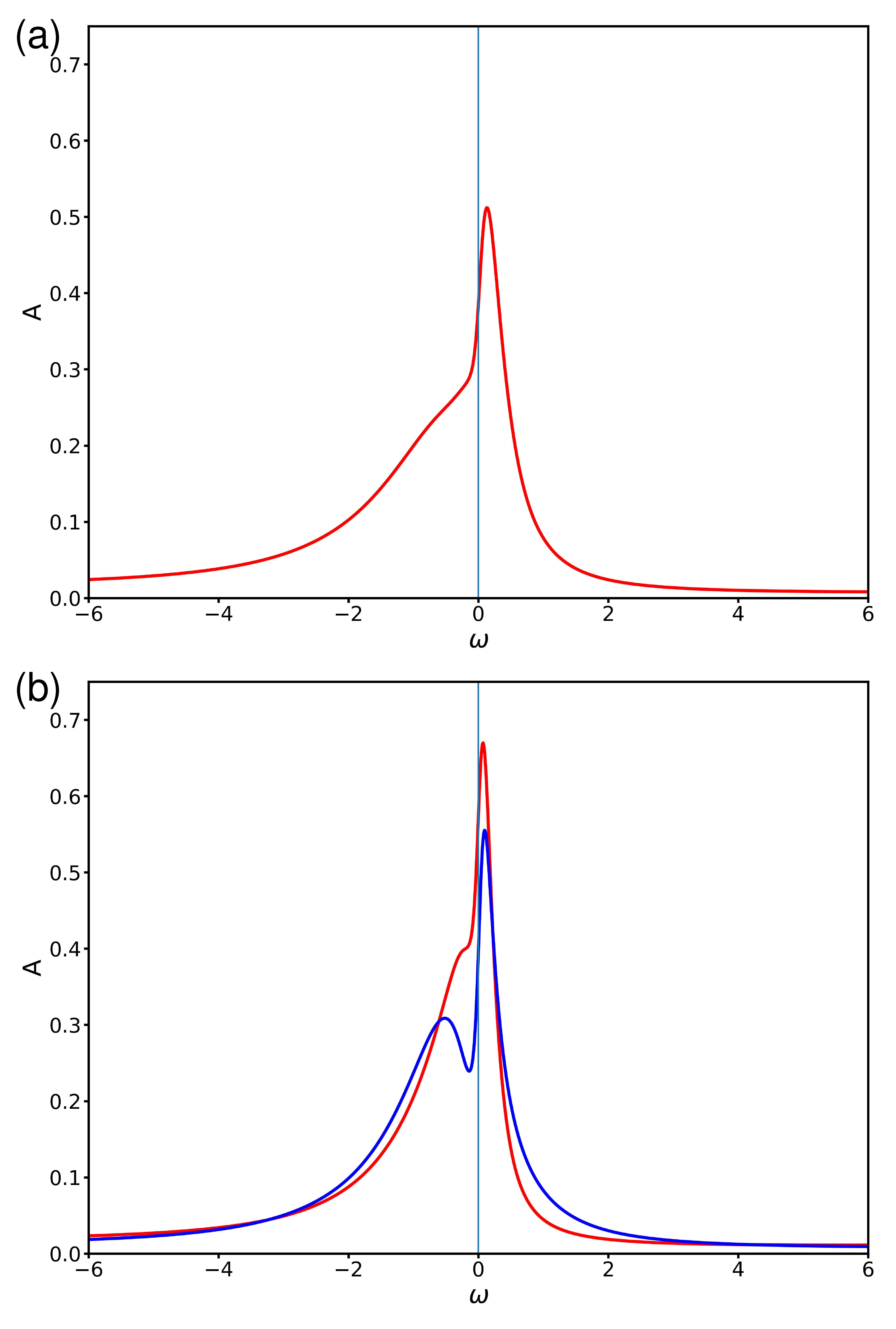}
    \caption{Correlated spectral functions of the 
    (a) cubic 3C and (b) hexagonal 4H phase of BRO, projected to three t$_{2g}$ states in the paramagnetic phase. The t$_{2g}$ states are degenerate in 3C phase
    while get split into doubly and singly degenerate states in 4H. }
    \label{para}
\end{figure}

The paramagnetic correlated spectral function for $U = 2.3$\,eV and $J = 0.4$\,eV, which are accepted values for ruthenate oxides,\cite{mravlje2011} in 3C and 4H phases at $\beta = 40$\,eV$^{-1}$
is shown in Fig.~\ref{para}. We see that in the paramagnetic phase both the structures have metallic ground states. In case of the 3C cubic state the three t$_{2g}$ orbitals are degenerate, while in case of 4H the degeneracy between the t$_{2g}$ orbitals is broken, with two degenerate orbitals and
another singly degenerate orbital arising due to the trigonal distortion of the RuO$_6$ octahedra in hexagonal symmetry.

\begin{figure}
    \centering
    \includegraphics[width=\columnwidth]{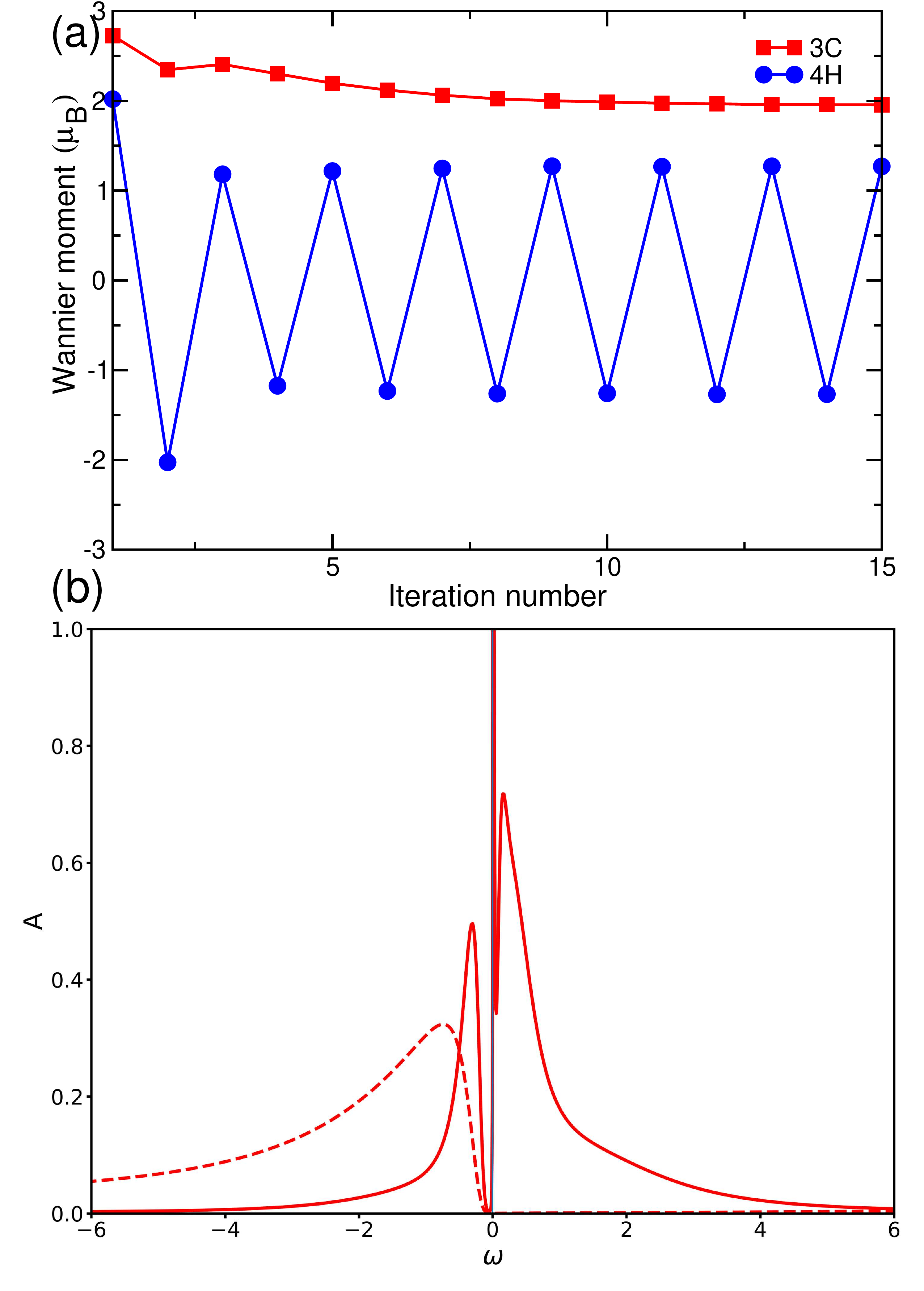}
    \caption{(a) Variation of Wannier moments as function of DMFT iterations for both 3C and 4H, showing stable FM moments for 3C and fluctuating moments for 4H. (b) Spectral function of the cubic 3C phase of BRO in the spin polarised phase. The dashed curve represents the up spin channel and the solid curve the down spin channel.} 
    \label{spin-split}
\end{figure}

Next, we proceed to exploring magnetism within single impurity DMFT by introducing
symmetry breaking. For that purpose, we start from the paramagnetic solutions,
add a symmetry breaking term in form of a spin splitting in the real part of
the self energies, and let the DMFT iterative
cycle converge to a possible symmetry-broken solution with net ordered magnetic
moment. We carry out the calculations at various different values of
inverse temperature with
$\beta$ between 40 and 200\,eV$^{-1}$, for both 3C and 4H structures. At $\beta=40$\,eV$^{-1}$, the calculations are found to converge to a
paramagnetic state, while upon reducing temperature, a transition to a magnetic solution is found. In Fig.~\ref{spin-split} (a), we show a plot of the ordered moments of Ru Wannier functions with the number of
 DMFT iterations. For the 3C phase we see a stable FM state. A critical temperature of $T_C\approx 116$\,K is determined for 3C, albeit the critical temperatures being
overestimated\cite{Jin2008} due to the mean field nature of the DMFT calculations. On the other hand, for 4H the ordered moments are found to be not stable, but they rather
fluctuate as a function
of iteration. The alternating or oscillating nature of moments within the single impurity DMFT hints towards propensity to antiferromagnetism, not being captured
within the single impurity DMFT.

We next take a look at the spectral functions for the spin-polarised DMFT calculations for the 3C structure, as shown in Fig.~\ref{spin-split} (b), calculated at $\beta=100$\,eV$^{-1}$, for $U=2.3$\,eV and $J=0.4$\,eV. We find that in case of 3C, a ferromagnetic correlated metallic state appears
with the up spin channel majorly occupied and the down spin channel majorly unoccupied. This is in agreement with the experimental observation.\cite{Jin2008}

\subsection{Variation of interaction parameters and influence on the magnetic phases}

Having established the importance of correlations in the description of properties of both 3C and 4H, we next explore
the effect of variation of the correlation strength, parametrized through parameters $U$ and $J$, on the magnetic properties of
3C and 4H. We carry out this exercise primarily due to two reasons. Although we will later in this section estimate the values of $U$ and $J$ form first-principles, it is well-known that this estimate carries some degree of uncertainty.\cite{crpa-werner} 
Second, it has been shown previously that it is possible to tune the Hubbard $U$ and Hund $J$ parameters by application of strain.\cite{crpa-franchini} We thus wish to study the trend of magnetism as a function of $U$ and $J$ values, highlighting the different trends in 3C and 4H, before we proceed to use first-principle estimates to place the actual materials in the phase diagram.

For this purpose, we repeat the symmetry-broken DMFT calculations at  $\beta=100$\,eV$^{-1}$, for a range of
$U$ and $J$ values, and monitor two quantities, Modulo $M$ and $M$.  Modulo $M$ refers to the absolute value of the ordered magnetic moment,
averaged over the last four iterations of the DMFT cycle, and $M$ is the magnetisation averaged
over the last four iterations including the sign of the magnetic moment. The quantity
 $M$ will have a value close to zero for moments fluctuating from plus to minus over iterations, and it will be equal to Modulo $M$ for stable ordered moments in a ferromagnetic state.

\begin{figure}
    \centering
    \includegraphics[width=\columnwidth]{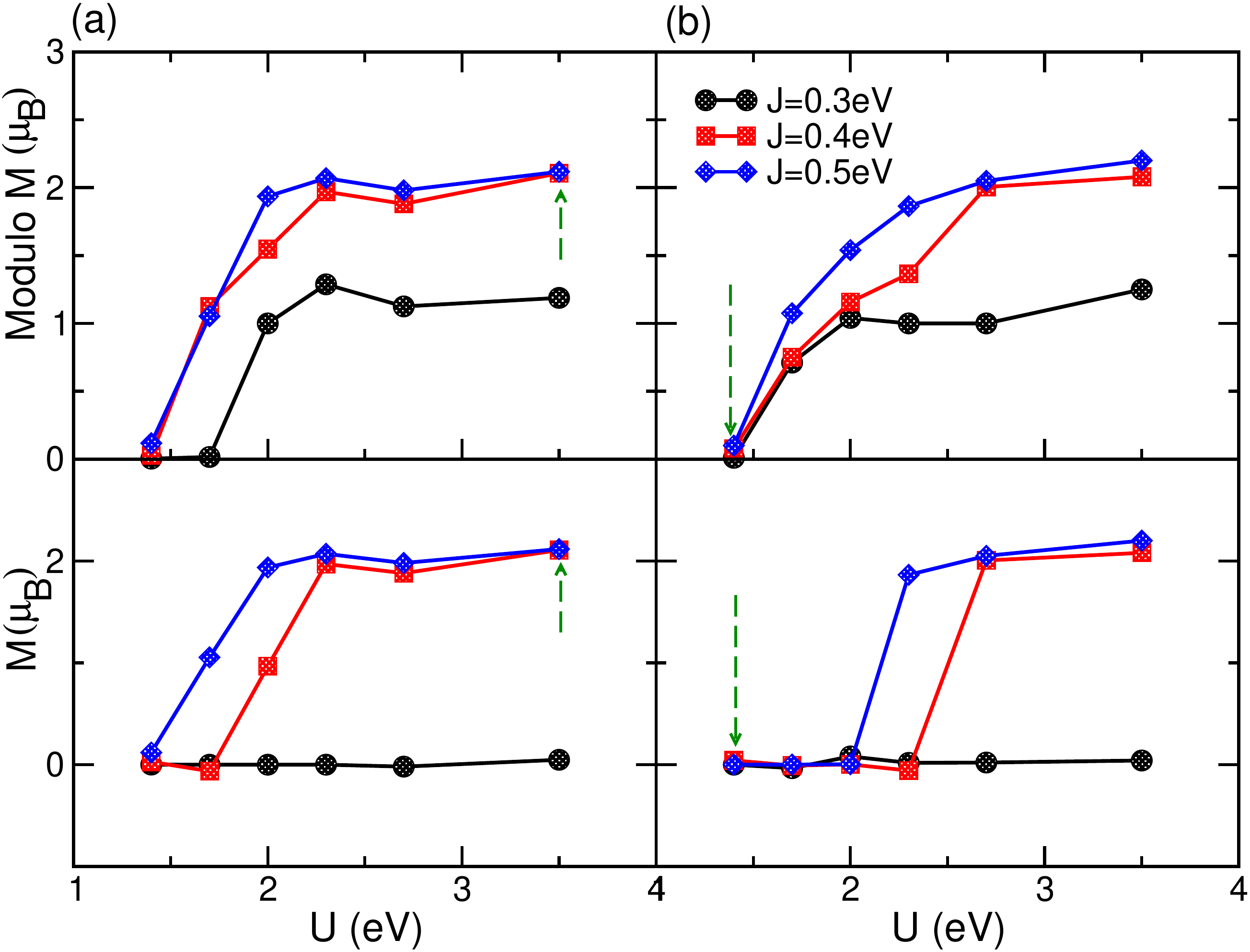}
    \caption{Variation of magnetic states for (a) cubic 3C and (b) hexagonal 4H phase with changes of $U$ and $J$. Modulo $M$ indicates the average value of absolute value of the magnetisation over the last 4 iterations, and $M$ indicates the average magnetisation over the last 4 iterations including its sign.
    The dashed green arrows in each case marks the cRPA values of $U$ and $J$.}
    \label{phase}
\end{figure}

The variation of Modulo $M$ and $M$ within a wide range of values of $U$ and $J$ is shown in Fig.~\ref{phase}.
A non-trivial variation of magnetic states is found to be achieved with variation of $U$ and $J$. One
can identify three regions in this phase diagram in general. At small values of $U$ and $J$ one finds
a paramagnetic state. With increasing $U$ but still small $J$, a region is found with moments fluctuating over iterations. 
With larger values of $U$ and $J$, a state with stable FM moments is seen. 

While this general feature is found to be
true for both 3C and 4H, there are subtle differences. For a fixed $U$ value of 2.3\,eV, 3C shows moments fluctuating over iterations
for $J= 0.3$\,eV, and stable FM moments for $J = 0.4-0.5$\,eV. The 4H phase, however, shows moments fluctuating over iterations for $J=0.3-0.4$\,eV, and
stable FM moments only for $J=0.5$\,eV. Fixing now $J=0.4$\,eV, upon varation of $U$, for the 3C structure a fluctuating
moment state is found for $U = 1.7$\,eV, while a FM state stabilizes beyond $U = 2.3$\,eV. On the other hand,
the 4H phase for the same $J = 0.4$\,eV shows fluctuating moments until $U = 2.3$\,eV, and stabilization of the FM state only beyond
$U = 2.7$\,eV. This exercise conclusively demonstrates that the magnetic state is crucially dependent on the choice
of $U$ and $J$, with critical $U$ and $J$ values for the stabilization of the FM state as compared to moments fluctuating over iterations being different between 3C and 4H.

The above exercise calls for the need of a first-principles evaluation of $U$ and $J$ values in 3C and 4H phases. As discussed above,
the change of Ru-O covalency due to differences in the connectivity of RuO$_6$ octahedra between 3C and 4H, is expected to
influence the screening and, thus, the value of $U$ and $J$. In particular, we carry out constrained RPA (cRPA) calculations as
implemented within VASP to calculate the Coulomb matrix elements $U_{{ijkl}}(\omega =0)$ for the Ru t$_{2g}$
states of BRO in both 4H and 3C phase. For 3C we obtain from cRPA screened $U=3.5$\,eV and $J=0.5$\,eV, and for 4H we obtain screened $U=1.4$\,eV and $J=0.3$\,eV. Even considering the standard errors in the estimation of $U$ and $J$ in the cRPA methods,
one can see that there is a significant difference between the interaction values for the two phases. The cRPA
estimated $U$ and $J$ values for 3C and 4H are indicated in Fig.~\ref{phase} with green dashed arrows. 
With the choice of cRPA estimates, 3C falls in the regime of a stable FM state as expected, while 4H falls in the regime
of fluctuating moments, with vanishingly small oscillating moments. For 4H, we further find, as the Hubbard $U$ is progressively decreased at fixed $J=0.3$\,eV, 
the ordered moment keeps decreasing from $U=2.3$ to $U=1.7$\,eV,
and finally vanishes at $U=1.4$\,eV. Thus, at the limit of the cRPA estimated value of Hubbard $U$ for 4H,
the ordered moment vanishes, indicating a lack of any tendency to magnetic long-range order.

\section{Multi-impurity dynamics}

To further elucidate on the nature of magnetism in 4H structure, we expand the unit cell
to include four Ru atoms in a supercell, and solve a four impurity problem in DMFT, with $U = 2$\,eV (slightly higher than cRPA estimate), and $J=0.3$\,eV. We carry out calculations at $\beta=80-120$\,eV$^{-1}$. The four impurity DMFT calculations at these temperatures result in metallic solutions
with vanishingly small ordered moments for 4H. However, the small moments show an antifferomagnetic orientation within the Ru-Ru dimer and a ferromagnetic orientation between the dimers. The vanishingly small ordered moment corroborates the
experimental finding of absence of long-range magnetic ordering at finite temperatures, with some antiferromagnetic fluctuations.\cite{Gulino1995} 
\begin{figure}
    \centering
    \includegraphics[width=\columnwidth]{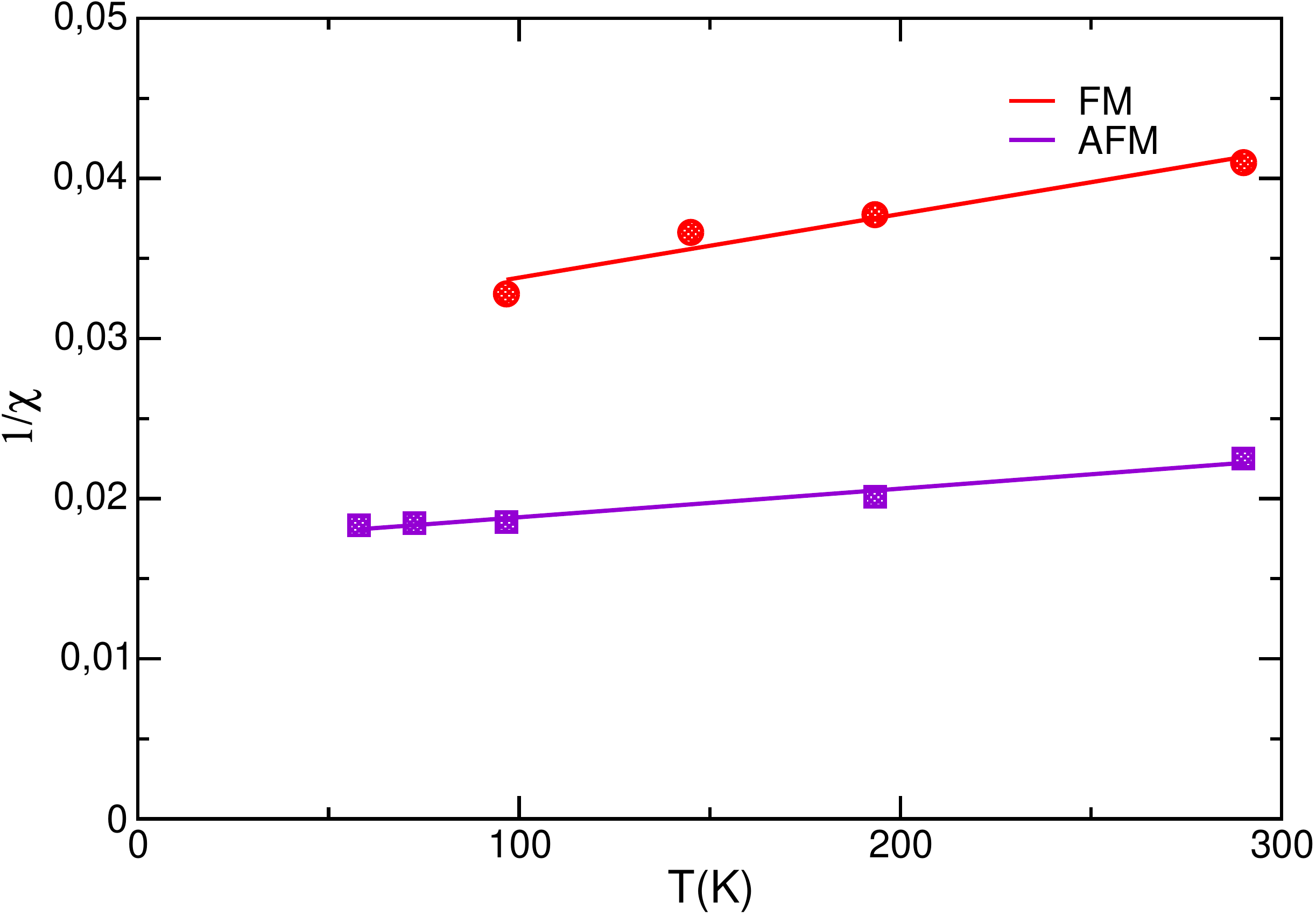}
    \caption{FM and AFM susceptibility for the 4H structure calculated in the supercell with four Ru atoms.}
    \label{spect-sus}
\end{figure}

To confirm this further, we carry out susceptibility calculations with both ferromagnetic and antiferromagnetic external magnetic fields (field pointing up on two impurities and pointing down on the other two impurities) on the 4H system . We vary the applied field from $0.01-0.05$\,eV
in steps of 0.01, and for each value of temperature we obtain the inverse slope of the magnetisation vs applied field within the linear regime.
This gives the inverse of the uniform susceptibility $1/\chi$ vs temperature $T$, as shown in  Fig.~\ref{spect-sus}.
By fitting the curve to a straight line in Fig.~\ref{spect-sus} we see that the antiferromagnetic susceptibility is much larger than the ferromagnetic susceptibility, thus confirming the possible presence of short-range antiferromagnetic fluctuations in the system. 

\subsection{Magnetic phase of 4H at zero temperature}

\begin{figure}
    \centering
    \includegraphics[width=\columnwidth]{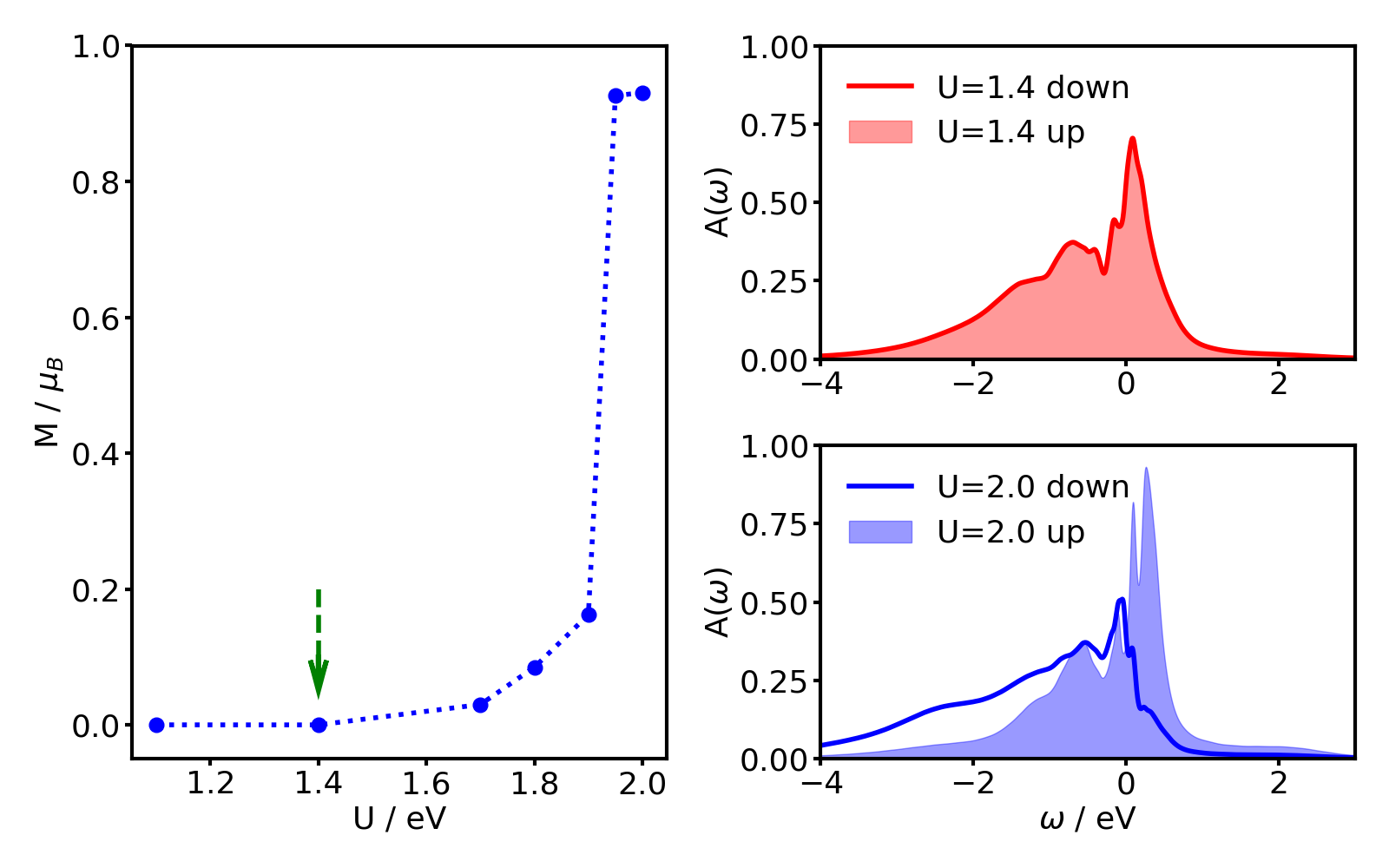}
    \caption{Wannier moments and density of states of 4H at $T=0$\,K as function of interaction strength $U$ (with constant $J=0.3$\,eV). The green arrow indicates the cRPA value of $U=1.4$\,eV. It can be seen that a small increase from the cRPA value is sufficient to lead to an (AFM) ordered state.}
    \label{ftps}
\end{figure}

Having established the absence of long-ranged magnetic ordering of 4H at finite temperature that were accessible by quantum Monte Carlo, we next investigate the magnetic phase at $T=0$\,K.
We carry out these DMFT calculations using the Fork Tensor Product States (FTPS) solver as implemented in TRIQS.
We do spin-polarised calculations with a density-density Hamiltonian, keeping $J=0.3$\,eV fixed and varying $U$ around the cRPA estimated
value, from $2$\,eV to $1.1$\,eV.

We find that at the larger values of $U=2$\,eV and $1.95$\,eV, the magnetic ordering is stabilized, leading to an antiferromagnetic state with the Ru atoms within the dimer aligned antiferromagnetically, and the dimers themselves being aligned ferromagnetically. It is noteworthy that this is the same antiferromagnetic state that
is found in spin-polarised DFT calculations in the same unit cell.
However, as the value of $U$ is decreased to $1.9$\,eV and further down towards the cRPA estimate of $1.4$\,eV, the ordered moments vanish after a few DMFT iterations.
This is shown in Fig.~\ref{ftps}. This leads to the conclusion that the magnetic state in 4H does not order in experimental studies due to
both thermal and quantum fluctuations. The spectral function calculated for $U = 2$\,eV and $U = 1.4$\,eV is also shown in Fig \ref{ftps}.
As is seen, the antiferromagnetic state at $U = 2$\,eV, and the nonmagnetic state at $U = 1.4$\,eV turn out to be metallic.

\section{Conclusion}

To conclude, considering the example of 3C and 4H polytypes of BaRuO$_3$ as test set, we investigate the effect of geometrical connectivity on magnetic properties of correlated transition metal oxides. The cubic 3C and hexagonal 4H phases with corner-shared versus face-shared connectivity of RuO$_6$ octahedra are reported to exhibit distinctly different magnetic behavior. While the 3C phase shows ferromagnetic ordering with moderately high magnetic transition temperature of $\sim 60$\,K, the 4H phase does not order magnetically, but shows rather paramagnetic behavior with evidence of short-range antiferromagnetic correlation.  

The single impurity DMFT calculations for a Ru t$_{2g}$ Hamiltonian shows a crossover from a generalised Fermi liquid to a non Fermi liquid kind of correlated behaviour upon variation of Hund’s coupling $J$ in both 3C and 4H phases, thus characterizing them as correlated Hund’s metals. Moreover, we find that the magnetism is highly dependent on the choice of Hubbard $U$ and Hund $J$ coupling, and the trend is not trivial. Depending on the choice of $U$ and $J$, either an ordered ferromagnetic state, or a paramagnetic state, or a state with ordered moments oscillating over DMFT iterations is achieved for both 3C and 4H. The ab-initio estimated $U$ and $J$ values through constrained RPA calculations yields significantly larger $U$ and $J$ values for the 3C phase as compared to the 4H phase. The ab-initio estimates of $U$ and $J$ place 3C in the FM region in the the $(U,J)$ parameter space of magnetic phases, while 4H is placed in a fluctuating magnetic state but with vanishingly small value of moment. Extending the DMFT calculations to the multi-impurity problem of four Ru atoms in a supercell shows that the magnetic state of 4H is indeed paramagnetic with antiferromagnetc short-range fluctuations. This is further confirmed by the uniform ferro- and antiferromagnetic susceptibilities, who show absence of long-range ordering but a larger AFM susceptibility. 
Finally, the FTPS calculations at $T=0$\,K show that the 4H phase is close to a long-range antiferromagnetically ordered metallic state, which can be stabilized upon slight increase of Hubbard $U$. This opens up the possibility of exploring exotic antiferromagnetic metallic phases in 4H BaRuO$_3$, by strain or a dielectric substrate which is expected to tune the screening, thus influencing the $U$ and $J$ values.

In summary, our study solves the puzzle of the contrasting magnetic behaviour of 4H and 3C polytypes of BaRuO$_3$, and provides a microscopic understanding in terms of the influence of geometric aspects on the magnetic behaviour of correlated oxides.

\begin{acknowledgments}
This work has been funded by the Austrian Science Fund (FWF): Y746. Calculations have partly been performed on the dcluster of TU Graz, and Vienna Scientific Cluster (VSC). TS-D acknowledges J.C.Bose National Fellowship (grant no. JCB/2020/000004) for funding.
\end{acknowledgments}

\bibliography{draft}

\end{document}